\documentstyle[12pt]{article}

 \newcommand{\be}{\begin{equation}}
 \newcommand{\ee}{\end{equation}}
 \newcommand{\ba}{\begin{eqnarray}}
 \newcommand{\ea}{\end{eqnarray}}
 
 \newcommand{\del}{\partial}

\newcommand{\lef}{\left}

\newcommand{\ri}{\right}

\newcommand{\ca}{{\cal A}}

\newcommand{\cf}{{\cal F}}

\newcommand{\cl}{{\cal L}}

\newcommand{\fr}{\frac}

\begin{document}

\begin{titlepage}

\topmargin -15mm

\vskip 10mm
\vskip 25mm

\centerline{ \LARGE\bf Quantized Skyrmion Fields }
\vskip 2mm
\centerline{ \LARGE\bf in 2+1 Dimensions}

    \vskip 2.0cm

    \centerline{\sc E.C.Marino }

     \vskip 0.6cm
     
\centerline{\it Instituto de F\'\i sica}
\centerline{\it Universidade Federal do Rio de Janeiro } 
\centerline{\it Cx.P. 68528, Rio de Janeiro, RJ 21945-970, Brazil} 
\vskip 2.0cm

\begin{abstract} 

A fully quantized field theory is developped
for the skyrmion topological excitations
of the O(3) symmetric CP$^1$-Nonlinear Sigma Model in 2+1D. The method
allows for the obtainment of arbitrary
correlation functions of quantum skyrmion
fields. The two-point function is evaluated in three different situations:
a) the pure theory; b) the case when it is coupled to fermions
which are otherwise non-interacting and c) the case when an 
electromagnetic interaction among the fermions is introduced.
The quantum skyrmion mass is explicitly obtained in each case from
the large distance behavior of the two-point function and the skyrmion
statistics is inferred from an analysis of the phase of this function.
The ratio between the quantum and classical skyrmion masses is obtained,
confirming the tendency, observed in semiclassical calculations, that
quantum effects will decrease the skyrmion mass.
A brief discussion of asymptotic skyrmion states, based on the short
distance behavior of the two-point function, is also presented.

\end{abstract}

\vskip 5mm

PACS numbers: 74.25.Ha, 11.27.+d, 75.25.+z

\vskip 20mm
Work supported in part by CNPq-Brazilian National Research Council.
     E-Mail address: marino@if.ufrj.br

\end{titlepage}

\hoffset= -10mm

\leftmargin 23mm

\topmargin -8mm
\hsize 153mm
 
\baselineskip 7mm
\setcounter{page}{2}

\section{Introduction}

\bigskip

Skyrmions are topologically nontrivial solutions which appear in the
O(3) symmetric Nonlinear Sigma Model (NSM) in 2+1D \cite{bp}. Sometimes they
are called ``baby skyrmions'', in order to distinguish them from their
3+1-dimensional counterparts. In the CP$^1$ version they become vortices
carrying a ``magnetic flux'' along the spatial plane. There is a great
interest in the study of the quantum properties of skyrmions because of
the central role played by the NSM in some important condensed matter
systems. A first example comprise the
systems presenting the quantum Hall effect.
A quantum field theory which neglects the spin of the electrons has been
proposed for the description of this effect \cite{zhk}. Later on the
spin was introduced in the theory
and this has been shown to be equivalent to
a NSM containing, as a consequence, skyrmion configurations \cite{skk}.
These skyrmions have been recently observed directly in neutron scattering
experiments performed in quantum Hall systems in the $\nu = 1$ level
\cite{sk}.

Another important class of planar condensed matter systems in which skyrmions
do play an important role are high-T$_c$ superconductors.
In the absence of doping, these materials are quite well described,
in the strong coupling limit, by a
NSM where the nonlinear sigma field is
the continuous version of the electron spin. The connection arises trough
the equivalence, on the continuum limit, between the 2+1-dimensional
O(3) NSM and the antiferromagnetic Heisenberg model in 2D \cite{ha}. This,
by its turn, provides a good description of the physics in the $CuO_2$
planes for the pure system in the strong coupling limit \cite{dk,ak}.
It has been shown that in the process of doping, formation of spin textures
or skyrmions accompany the introduction of holes in the $CuO_2$-planes
\cite{st,rod,go,sh,emsc}. These eventually lead to the destruction of the
N\'eel ordering which is known to exist at low doping. Other very
interesting effects due to skyrmions, like the modification of both the
the NMR line shapes and the magnetic structure factor, the latter affecting
neutron scattering cross sections, have been predicted for skyrmions
introduced by doping antiferromagnetic planar systems \cite{st}.

The above examples justify the importance of having a well established and
manageable quantum theory of skyrmions in the 2+1-dimensional
CP$^1$-Nonlinear Sigma Model. A semiclassical effective lagrangian describing
the skyrmion physics was derived in \cite{st,pw}.
The question of quantum effects on skyrmions was addressed in \cite{rod},
where a semiclassical quantization of skyrmions was developped 
allowing for the obtainment of spin correlators in the presence of skyrmions,
as well as the effective interaction energy of quantum skyrmions.
An attempt to obtain a quantum theory of skyrmion fields was made in
\cite{fm}. The skyrmion correlation functions, however, were not evaluated
there. In the present work, we obtain a full quantum theory for the skyrmion
field within the CP$^1$ version of the theory, taking advantage of the fact
that in this framework the skyrmions are vortices and the theory of
quantum vortices developped in \cite{nv,vor} can be therefore applied. The
quantum skyrmion two-point correlation function is explicitly evaluated
in three different situations. Firstly in the pure CP$^1$-NSM. Secondly,
when Dirac fermions are minimally coupled to the CP$^1$ gauge field but
their mutual direct interaction is neglected.
Finally, when the electromagnetic interaction among the fermions is included.
In this latter case the true, three-dimensional,
electromagnetic interaction is considered, even
though, the fermions are moving on a plane \cite{enp}. In all of these cases,
we explicitly obtain the quantum skyrmion mass through the large distance
behavior of the two-point function. This is compared to the classical mass
and the tendency of quantum corrections to decrease the skyrmion mass
observed semiclassically in \cite{rod}, is confirmed.
The skyrmion statistics is also evaluated in each case from the behavior
of the phase of this function. The explicit results show that the
conjecture, formulated in \cite{fm}, about the condensation of skyrmions
in the absence of a Chern-Simons term is actually not valid.
A brief discussion about the existence of asymptotic quantum skyrmion states
is presented at the conclusion.

\section{The CP$^1$-Nonlinear Sigma Model and a Quantum Field
Theory of Skyrmions}

\setcounter{equation}{0}
\bigskip

Let us consider the O(3) Nonlinear Sigma Model which describes a field
$n^a, a=1,2,3$ subject to the constraint $n^a n^a = \rho_0^2$. The lagrangian 
is simply
\be
\cl = \fr{1}{2} \del_\mu n^a \del^\mu n^a
\label{1}
\ee
and therefore the whole dynamics comes from the constraint. The great
importance of this model for condensed matter systems emerges from the
fact that it is the continuum limit of
the O(3)-symmetric antiferromagnetic Heisenberg
model in a two-dimensional square lattice \cite{ha}, which is described
by the hamiltonian
\be
H = J \sum_{i,j} \vec S_i \cdot \vec S_j
\label{1a}
\ee
with $J > 0$. The nonlinear sigma field $ \vec n$ is the continuum limit
of the staggered spin corresponding to $\vec S$ and $\rho_0^2$ satisfies
\cite{ha} $\fr{\hbar c}{a} = \rho_0^2  \fr{2 \sqrt{2}}{\sqrt{S(S+1)}}$,
$S$ being the spin quantum number, $c$ the spin-wave velocity
and $a$, the lattice spacing. Using linear spin-wave theory
results for the Heisenberg model, on the other hand, it is found that
\cite{ak} $\fr{\hbar c}{a} = 1.18 (2S) \sqrt{2} J$
where $J$ is the Heisenberg antiferromagnetic coupling constant.
These two relations
are valid for large $S$. Assuming the ratio of both holds for any $S$,
we establish the following relation between the coupling constants in
the continuum and in the lattice, for the case $S=1/2$
\be
\rho_0^2 = 1.18 \fr{\sqrt{3}}{4} J.
\label{1b}
\ee

It is very
convenient to express the $n^a$-field in the so called CP$^1$ language,
in terms of a doublet of complex scalar fields $z_i, i=1,2$, which also
satisfy the constraint $|z_1|^2 + |z_2|^2 = \rho_0^2 $. The mapping
between the two fields is established through
\be
n^a = \fr{1}{\rho_0} z_i^\dagger \sigma^a_{ij} z_j
\label{2}
\ee
where $\sigma^a$ are Pauli matrices.
The corresponding lagrangian in the CP$^1$ version is \cite{pol,raj}
\be
\cl = 2 \ (D_\mu z_i)^\dagger (D^\mu z_i)
\label{3}
\ee
where $D_\mu = \del_\mu + i A_\mu$ and 
$A_\mu = \fr{i}{\rho_0^2} z_i^\dagger \del_\mu z_i$.
Notice the well known spontaneous
generation of a local U(1) symmetry in this version.

It is convenient to express the $z_i$-fields in the polar representation
\be
z_i = \fr{\rho_i}{\sqrt{2}} e^{i \theta_i}
\label{4}
\ee
where $\rho_i$, $\theta_i$ are real fields. The CP$^1 $ constraint
becomes then
\be
\fr{1}{2}\lef ( \rho_1^2 + \rho_2^2 \ri ) = \rho_0^2
\label{5}
\ee
Inserting (\ref{4}) in the lagrangian (\ref{3}), we get
\be
\cl =  \sum_{i=1}^2 \lef \{ \del_\mu \rho_i \del^\mu \rho_i +
\rho_i ^2 \lef [ A_\mu + \del_\mu \theta_i \ri  ]^2 \ri  \}
\label{6}
\ee
The gauge transformation under which (\ref{6}) is invariant is
$$
\theta_i \rightarrow \theta_i + \Lambda (x)
$$
\be
A_\mu \rightarrow A_\mu - \del_\mu \Lambda (x)
\label{7}
\ee
Let us introduce now an explicitly gauge invariant description through
the field
\be
\chi_i = \theta_i - \fr{ \del_\alpha A^\alpha}{(-\Box)}
\label{8}
\ee
which is invariant under (\ref{7}).
Introducing (\ref{8}) in (\ref{6}), we get
\be
\cl =  \sum_{i=1}^2 \lef \{ \del_\mu \rho_i \del^\mu \rho_i +
\rho_i ^2 \lef [ \fr{ \del_\alpha F^{\mu\alpha}}{(-\Box)}
+ \del_\mu \chi_i \ri  ]^2 \ri \}
\label{9}
\ee
which is explicitly gauge invariant.
The euclidean action corresponding to this can be rewritten, after some
algebra \cite{nv} as
$$
S = \int d^3z \lef \{ 
\sum_{i=1}^2 \lef [ \del_\mu \rho_i \del^\mu \rho_i +
\rho_i^2 \del_\mu \chi_i \del^\mu \chi_i -
\del_\mu \lef ( \rho_i^2 \ri )
\lef [ \fr{ \del_\alpha F^{\mu\alpha}}{(-\Box)} \ri ] \chi_i \ri ]
+ \ri .
$$
\be
\lef . 
+ \fr{1}{4} F_{\mu\nu} \lef [ \fr{2 (\rho_1^2 + \rho_2^2)}{(-\-\Box)} \ri ]
F^{\mu\nu}\ri \}
\label{10}
\ee

Before proceeding, let us make some remarks about the topological aspects
of the CP$^1$-Nonlinear Sigma Model. There are two nontrivial topologies
associated to the mappings $S^2 \rightarrow S^2$ and $S^3 \rightarrow S^2$,
which are carried on, respectively, by static and non-static configurations
of the $n^a$-field which, because of the constraint, lives on an $S^2$
manifold. The corresponding topological invariants are, respectively the
topological charge
\be
Q= \int d^2x J^0
\label{11}
\ee
and the Hopf invariant
\be
S_H = \int d^3 x J^\mu A_\mu
\label{12}
\ee
In the two previous expressions, $J^\mu$ is the topological current, which
is given, respectively, in the Nonlinear Sigma and CP$^1$ versions as
\be
J^\mu = \fr{1}{8\pi} \epsilon^{\mu\alpha\beta} \epsilon^{abc} n^a
\del_\alpha n^b \del_\beta n^c
\label{13}
\ee
and
\be
J^\mu  = \fr{1}{2\pi} \epsilon^{\mu\alpha\beta} \del_\alpha A_\beta
\label{14}
\ee

The immediate consequence of the existence of the nontrivial topology in
the mapping $S^3 \rightarrow S^2$ is that, when performing functional
integrations over the basic fields, we must weigh the corresponding
topological sector by $e^{i \theta S_H}$. This implies that a term
$\theta S_H$ must be added to the action in (\ref{10}) \cite{ss}. In the
CP$^1$ language, this is nothing but a Chern-Simons term.

As a consequence of the nontriviality of the mapping $S^2 \rightarrow S^2$,
on the other hand, the theory possesses classical soliton solutions.
These are called ``skyrmions'' and have  topological charge $Q=1$.
In the Nonlinear Sigma version
the classical skyrmion solution is given by \cite{bp}
\be
\vec n_S (\vec x, 0) = \rho_0 \lef ( \sin f(r) \hat r, \cos f(r) \ri )
\label{15}
\ee
with
$$
f(r) = 2 \arctan \fr{\lambda}{r}
$$
where $\lambda $ is an arbitrary scale and $r$ is the radial distance in
two-dimensional space. In the CP$^1$ language, the skyrmion becomes a vortex.
This can be already inferred from (\ref{11}) and (\ref{14}) which show
that in this language the topological charge is the magnetic flux of the
field $A_\mu$ along the two-dimensional space. Indeed, using (\ref{2})
and (\ref{15}), we get
\be
z^a_S(\vec x, 0) = \rho_0
 \lef (  \begin{array}{c}
 \cos \fr{f(r)}{2} \exp^{- \fr{i}{2} arg(\vec r)}  \\
 \sin \fr{f(r)}{2} \exp^{ \fr{i}{2} arg(\vec r)} 
          \end{array} \ri )  
\label{16}
\ee
and
\be
\vec A_i^S = \fr{1}{2} \cos f(r) \del_i arg (\vec r)\ \ ,\ \  A_0 =0
\label{17}
\ee

From the explicit form of the above classical solution, we can
evaluate the classical skyrmion mass (energy) which derives directly from
(\ref{1}), namely,
\be
{\cal M}_{cl} = \fr{1}{2} \int d^2 x \  \vec \nabla n^a \cdot
\vec \nabla n^a
\label{177}
\ee
obtaining ${\cal M}_{cl} = 4 \pi \rho_0^2$ (see also \cite{bp,rod}).

A full quantum theory of vortices has been developed based on the idea of
order-disorder duality \cite{nv,vor} and therefore we can apply it in
the present case, in order to obtain a quantum field theory of skyrmions in
the CP$^1$ language. For a theory containing a vector field $A_\mu$, the
quantum skyrmion field is given by \cite{nv}
\be
\mu (\vec x, t) = \exp \lef \{ i\  2\pi \int_{\vec x,L}^\infty d \xi^i
\epsilon^{ij} \Pi^j (\vec \xi, t) \ri \}
\label{18}
\ee
where $\Pi^i$ is the momentum canonically conjugate to $A^i$. It follows that
\cite{nv}
\be
[Q, \mu(x)] = \mu(x)
\label{19}
\ee
and
\be
[\mu(x), A_i(y)] = \mu(x) \del_i^{(y)} arg(\vec y - \vec x)
\label{20}
\ee
These two equations show that indeed $\mu(x)$ creates a quantum vortex at
$x$ and the state $|\mu>$ is an eigenstate of the topological charge $Q$
with eigenvalue equal to one. It follows from the formalism developped in
\cite{nv,vor} that a path independent functional integral
describing the euclidean correlation functions of the vortex
operator can be obtained by adding to the field intensity tensor $F^{\mu\nu}$
an external field
\be
\tilde B^{\mu\nu} = b \int_{\vec x,L}^\infty d \xi_\lambda
\epsilon^{\lambda \mu \nu} \delta^3(z-\xi)
\label{21}
\ee
where $\fr{b}{2\pi}$ is the number of units of magnetic flux (topological
charge) or vorticity. For the unit skyrmion, $b=2\pi$.
Notice that (\ref{21}) is the magnetic field intensity (magnetic flux/
topological charge density) of a vortex with universe line along $L$.
For the CP$^1$ theory, described by (\ref{10}), the local
(path independent) euclidean skyrmion correlation function,
including the topological $\theta$-term is given by
\cite{nv}
$$
<\mu(x) \mu^\dagger(y)> = Z_0^{-1} \int  D\rho_i  \rho_i D\chi_i D A_\mu
\exp \lef \{ - \int d^3 z \lef [
 \sum_{i=1}^2 \lef [ \del_\mu \rho_i \del^\mu \rho_i +
\rho_i^2 \del_\mu \chi_i \del^\mu \chi_i + \ri.\ri.\ri.
$$
$$
\lef.
\del_\mu \lef ( \rho_i^2 \ri )
\lef [ \fr{ \del_\alpha \lef (F^{\mu\alpha}+ \tilde B^{\mu\alpha} \ri )}
{(-\Box)} \ri ] \chi_i \ri ] + 
\fr{1}{4} \lef ( F_{\mu\nu}+ \tilde B^{\mu\nu}\ri )
\lef [ \fr{2 ( \rho_1^2 + \rho_2^2 )}{(-\Box)} \ri ]
\lef ( F^{\mu\nu}+ \tilde B^{\mu\nu}\ri )
$$
\be 
\lef.\lef.
- \fr{i \theta}{4} \epsilon^{\mu\alpha\beta}
\lef [ \fr{\del_\nu ( F^{\nu\mu} + \tilde B^{\nu\mu})}{(-\Box)} \ri ]
\lef ( F_{\alpha\beta} + \tilde B_{\alpha\beta} \ri )
\ri ] \ri \}
\label{22}
\ee
In the above expression, $\tilde B^{\mu\nu}= \tilde B^{\mu\nu}(z;x) -
\tilde B^{\mu\nu}(z;y)$, corresponding to the field operators $\mu(x)$ and
$\mu^\dagger(x)$, respectively. Notice that we rewrote the Chern-Simons
term in terms of $F^{\mu\nu}$, up to a total derivative. The above
expression for the skyrmion correlation function can be easily shown
to be local (path independent), by performing the change in the functional
integration variable \cite{nv}
\be
A_\mu \longrightarrow A_\mu'= A_\mu + \Omega_\mu \ \   ; 
\ \  D A_\mu =D A_\mu'
\label{23}
\ee
with
\be
\Omega_\mu = b \int_{S(L,L')} d^2\xi_\mu \delta^3 (z-\xi)
\label{24}
\ee
where $S(L,L')$ is the surface closed by the curve $L$ appearing in
(\ref{21}) and another arbitrary curve $L'$. It is easy to show \cite{nv}
that under (\ref{23}),
\be
\tilde F^{\mu\nu} \longrightarrow
\tilde F^{\mu\nu} + \tilde B^{\mu\nu}(L') - \tilde B^{\mu\nu}(L)
\label{25}
\ee
Inspection of (\ref{22}), then, immediately shows its independence on $L$.
Arbitrary higher point correlation functions can be obtained by just
inserting additional external fields $\tilde B_{\mu\nu}$ in (\ref{22}).

\section{The Skyrmion Correlation Function}

\setcounter{equation}{0}
\bigskip

In this section we explicitly evaluate the two-point skyrmion correlation
function, Eq. (\ref{22}). In order to perform the integration in (\ref{22}),
we are going to use the saddle-point approximation.
To implement this, we introduce the constraint through a Lagrange
multiplier field, thereby obtaining (\ref{5}) as an equation of motion. 
Choosing a stationary solution where the moduli of the complex
scalar fields $z_i$ are taken to be constants,
namely, $\rho_i^2 = \rho_{i,0}^2$, amounts to selecting a given direction
in the manifold of minima
$\rho_{1,0}^2 + \rho_{2,0}^2 = 2 \rho_0^2$. We are going
to expand the $\rho_i$-fields around this constant solution. This
approximation was used for the evaluation of vortex correlation
functions in the relativistic Landau-Ginzburg theory in 2+1 and in 3+1
dimensions \cite{nv,qms}. Note that this approximation does not impose a
restriction on the topological charge, which is determined by the phases
of the $z_i$-fields, according to (\ref{16}) and (\ref{17}). Also, it goes
beyond the
semiclassical approximation for the skyrmion correlation function, because
the quantized skyrmion operators already include full quantum effects.

In this approximation, the
functional integral (\ref{22}) becomes, after integration over $\rho$
and $\chi$,
$$
<\mu(x) \mu^\dagger(y)> = Z_0^{-1} \int D A_\mu
\exp \lef \{ - \int d^3 z \lef [
\fr{1}{4} \lef ( F_{\mu\nu}+ \tilde B^{\mu\nu}\ri )
\lef [ \fr{4 \rho_0^2 }{(-\-\Box)} \ri ]
\lef ( F^{\mu\nu}+ \tilde B^{\mu\nu}\ri ) 
+ \ri.\ri.
$$
\be
\lef.\lef.
- \fr{i \theta}{4} \epsilon^{\mu\alpha\beta}
\lef [ \fr{\del_\nu ( F^{\nu\mu} + \tilde B^{\nu\mu})}{(-\Box)} \ri ]
\lef ( F_{\alpha\beta} + \tilde B_{\alpha\beta} \ri )
+ \fr{\xi}{2} (\del_\mu A^\mu)^2
\ri ] \ri \}
\label{26}
\ee
where the last term is a gauge fixing which has been inserted.
The above functional integral is quadratic and can be evaluated
by using the euclidean propagator of the $A_\mu$ field, which is given
in momentum space by
\be
D^{\mu\nu} (k) =  \fr{1}{[M^2 + \theta^2 k^2]} \lef [ \fr{M}{k^2}
\lef ( k^2 \delta^{\mu\nu} - k^\mu k^\nu \ri )
- \theta \epsilon^{\mu\lambda\nu} k_\lambda \ri ] +
\lef [ \fr{[M^2 - \theta^2 k^2]}{[M^2 + \theta^2 k^2]} \ri ]
\fr{k^\mu k^\nu}{\xi k^4}
\label{27}
\ee
where $M = 4 \rho_0^2$.
Inserting (\ref{21}) in (\ref{26}), we see that
the linear term in $A_\mu$ in (\ref{26}) can be written
as
\be
\int d^3z L^\nu (z;x,y) A_\nu
\label{28}
\ee
where
$$
L^\nu (z;x,y) = \tilde B^\nu (z;x,y) + \tilde C^\nu (z;x,y)
$$
with
$$
\tilde B^\nu (z;x,y) = M b \int_{x,L}^y d\xi_\alpha
\epsilon^{\alpha\beta\nu} \del_\beta \lef [ \fr{1}{-\Box} \ri ] (z-\xi)
$$
and
\be
\tilde C^\nu (z;x,y) = i \theta b \int_{x,L}^y d\xi_\mu
\lef (-\Box \delta^{\mu\nu} +\del^\mu \del^\nu \ri )
\lef [ \fr{1}{-\Box} \ri ] (z-\xi)
\label{29}
\ee
Integration over $A_\mu$ in (\ref{26}) gives
\be
<\mu(x) \mu^\dagger(y)> = 
\exp \lef \{ \fr{1}{2} \int d^3 z d^3 z'  L^\mu (z;x,y) L^\nu (z';x,y)
D^{\mu\nu} (z-z') - S_L \ri \}
\label{30}
\ee
where $S_L$ is the line dependent renormalization factors of (\ref{26})
which ensure the locality of $<\mu \mu^\dagger>$:
\be
S_L = S_{1,L} + S_{2,L} =
\fr{1}{4} \int d^3 z  \tilde B^{\mu\nu} \lef [ \fr{M}{(-\-\Box)} \ri ]
\tilde B^{\mu\nu} +
\fr{i \theta}{4}  \int d^3 z 
\epsilon^{\mu\alpha\beta}
\lef [ \fr{\del_\nu \tilde B^{\nu\mu}}{(-\Box)} \ri ]
\tilde B_{\alpha\beta} 
\label{31}
\ee
From the form of (\ref{29}) we can see that only the
gauge independent first two terms of
(\ref{27}) contribute to (\ref{30}), thereby
ensuring the gauge independence of the skyrmion correlation function.

There are six terms coming from the first part of the exponent in
(\ref{30}), which in an obvious notation we call $BMB, CMC, B\theta C,
BMC, B \theta B, C \theta C$. The first three ones combine in the form
\be
T_1 = - \fr{b^2}{2} \int_{x,L}^y d \xi_\mu \int_{x,L}^y d \eta_\nu
\lef [ - \Box \delta^{\mu\nu} + \del^\mu \del^\nu \ri ]
\lef [ \fr{M}{\epsilon}
- \fr{|\xi - \eta|}{8 \pi} \ri ]
\label{32}
\ee
where the last expression between brackets is
$\lef [ \fr{M}{(-\Box)^2} \ri ] \equiv \cf^{-1} \lef [ \fr{M}{k^4} \ri ] $
and $\epsilon$ is a regulator for the
Fourier transform of $\fr{1}{k^4}$.
The last three terms combine in the form
\be
T_2 = i \fr{\theta b^2}{2} \int_{x,L}^y d \xi_\mu \int_{x,L}^y d \eta_\nu
\epsilon^{\mu\lambda\nu} \del_\lambda  \lef [ 
\fr{1}{4 \pi |\xi - \eta|} \ri ] =
i \fr{\theta b^2}{4 \pi} \lef [ arg (\vec x -\vec y) + arg (\vec y -\vec x)
\ri ]
\label{33}
\ee
where the expression between brackets in the first part is
$\lef [ \fr{1}{(-\Box)} \ri ] \equiv \cf^{-1} \lef [ \fr{1}{k^2} \ri ]$
and the second equality is proved in
\cite{cmfc}.

Using (\ref{21}) and integrating the second term in (\ref{32}), we get
\be
T_1 = - \fr{M b^2}{8 \pi} |x -y|
+ \fr{1}{4} \int d^3z \tilde B^{\mu\nu}
\lef [ \fr{M}{- \Box}  \ri ] \tilde B^{\mu\nu}
\label{331}
\ee
Inserting (\ref{21}) in (\ref{31}), we see that $S_{2,L}$ cancels $T_2$.
Both terms are actually multivalued, as can be seen from (\ref{33}).
This is related to the statistics of skyrmions. We shall return to this
point in Sect. 6.
Also $S_{1,L}$ cancels the last, line dependent term, of $T_1$ when all
the terms are put together in (\ref{30}). We finally conclude that only
the first term of (\ref{331}) remains in the exponent of (\ref{30}) and
\be
<\mu(x) \mu^\dagger(y)> = 
\exp \lef \{- \fr{ \rho_0^2 b^2}{2 \pi} |x -y| \ri \}
\label{34}
\ee
where we have used $M = 4 \rho_0^2$. This is our result for the skyrmion
euclidean two-point function. From it we infer that the quantum skyrmion mass
is ${\cal M} = \fr{ \rho_0^2 b^2}{2 \pi}$. Observe that for the unit
skyrmion, $b =2\pi$ and ${\cal M} = 2 \pi \rho_0^2$. This is to be compared
with the classical skyrmion mass calculated in (\ref{177}), namely
${\cal M}_{cl} =  4 \pi \rho_0^2$. The ratio ${\cal M}/{\cal M}_{cl} = 0.5$
confirms the fact observed semiclassically \cite{rod} that quantum
corrections decrease the skyrmion mass. This can be understood on general
grounds by examining the expression for the classical energy given either
by (\ref{177}) or (\ref{1a}). From the latter we can infer that the
classical energy must be proportional to
the spin stiffness $\rho_s = S^2 J$. It is reasonable to expect that in the
semiclassical approximation, it must be proportional to the
renormalized spin stiffness $\rho_s = Z_\rho S^2 J$ \cite{ak}. Since
\\ $Z_\rho \simeq 0.7$ \cite{ak}, we should expect semiclassically that
${\cal M}/{\cal M}_{cl} \simeq 0.7$. Our result for the skyrmion mass,
however, is based on the analysis of the large
distance decay of the fully quantized
skyrmion operator correlation function and includes full quantum effects,
going beyond the semiclassical approach were expressions
(\ref{177}) or (\ref{1a}) are used for obtaining the skyrmion energy.

\section{The Introduction of Fermions}

\setcounter{equation}{0}
\bigskip

Let us investigate in this section how the coupling of fermions influences
the skyrmion properties at the quantum level. For this, let us consider the
lagrangian
\be
\cl = 2\  (D_\mu z_i)^\dagger (D^\mu z_i) + i \fr{\theta}{2}
\epsilon^{\mu\alpha\beta} A_\mu \del_\alpha A_\beta +
i \bar \psi \not\! \del \psi
- m \bar \psi \psi - q \bar \psi \gamma^\mu \psi A_\mu
\label{35}
\ee
where $q$ is the coupling between the fermions and the $CP^1$ gauge field.
We can obtain an effective theory for $A_\mu$ by integrating over the
fermions. This will be done in two limits of approximation, namely, large
and small fermion mass $m$, both for small $q$. In this case, we can
write the fermionic determinant as \cite{fd},
\be
\ln Det \lef [ i \not\! \del + \not\!\! A - m \ri ] =
\fr{q^2}{2} \int d^3 z \lef [ \fr{A}{2}
\epsilon^{\mu\alpha\beta} A_\mu \del_\alpha A_\beta +
\fr{1}{4} F_{\mu\nu} \lef [ B \ri ] F^{\mu\nu} \ri ]
\label{36}
\ee

Let us consider the large mass limit first. In this case \cite{fd},
$$
A = \fr{q^2}{2 \pi} + O \lef (\fr{1}{m^2} \ri )
$$
and
\be
B = \fr{q^2}{16 \pi m} + O \lef (\fr{1}{m^3} \ri )
\label{37}
\ee
The result of the fermion integration can be written as the exponential of
(\ref{36}). Combining this with the result of the integration over
the $CP^1$ fields, which can be done in the same approximation as in the
previous section, we arrive at an expression for the
skyrmion correlation function, which is identical to (\ref{26}),
except for the following modifications
$$
\theta \longrightarrow \theta + \fr{q^2}{2 \pi}
$$
\be
M \longrightarrow M + \fr{q^2(-\Box)}{16 \pi m}
\label{38}
\ee
the last one being $M \rightarrow M + \fr{q^2 k^2}{16 \pi m}$ in
momentum space. Substituting these modififications in (\ref{27}), (\ref{29})
and thereafter, we arrive at an expression like (\ref{32}) but with
the last term between brackets being now modified as
\be
\lef [ \fr{M}{(-\Box)^2} \ri ] \longrightarrow
\lef [ \fr{M}{(-\Box)^2} + \fr{q^2}{16 \pi m (-\Box)} \ri ]
\label{39}
\ee
The nonlocal terms cancel exactly as before. Going through the same steps
as in the preceeding section, after (\ref{32})
and using the fact that
\be
\lef [ \fr{1}{-\Box} \ri ] \equiv
{\cal F}^{-1}\lef [ \fr{1}{k^2} \ri ] = \fr{1}{4 \pi |x - y|}  
\label{391}
\ee
we arrive at
\be
<\mu(x) \mu^\dagger(y)>_{LM} = 
\exp \lef \{- \fr{ \rho_0^2 b^2}{2 \pi} |x - y| +
\fr{q^2 b^2}{64 \pi^2 m |x - y|} \ri \}
\label{40}
\ee
This is the quantum skyrmion correlation function in the presence of 
fermions with a large mass $m$ and coupling constant $q$. Observe that the
skyrmion mass, determined by the large distance behavior
of (\ref{40}), is not modified by the presence of fermions. The short
distance behavior of the two-point function, however, is completely
different, being highly singular in this case.

Let us turn now to the small mass limit. In this case \cite{fd},
$$
A = \fr{q^2}{4 \pi} + O(m)
$$
and
\be
B(k) = \fr{q^2}{16 k} + O(m)
\label{41}
\ee
Now, the modifications corresponding to (\ref{38}) are
$$
\theta \longrightarrow \theta + \fr{q^2}{4 \pi}
$$
\be
M \longrightarrow M + \fr{q^2(-\Box)^{1/2}}{16}
\label{42}
\ee
and the expression corresponding to (\ref{39}) is modified as
\be
\lef [ \fr{M}{(-\Box)^2} \ri ] \longrightarrow
\lef [ \fr{M}{(-\Box)^2} + \fr{q^2}{16 (-\Box)^{3/2}} \ri ]
\label{43}
\ee
Repeating the same procedure as in the last section, after (\ref{32})
and using the fact that
\be
\lef [ \fr{1}{(-\Box)^{3/2}} \ri ] \equiv
{\cal F}^{-1}\lef [ \fr{1}{k^3} \ri ] = -\fr{1}{2 \pi} \ln |x - y|
\label{433}
\ee
we get
\be
<\mu(x) \mu^\dagger(y)>_{SM} = 
\exp \lef \{- \fr{ \rho_0^2 b^2}{2 \pi} |x - y| -
\fr{q^2 b^2}{ 32 \pi^2} \ln |x - y| \ri \} =
\fr{\exp \lef \{- \fr{\rho_0^2 b^2}{4 \pi} |x - y|
\ri \}}{|x - y|^{\fr{q^2 b^2}{ 32 \pi^2}}}
\label{44}
\ee
This is the skyrmion correlation function in the small fermion mass limit
(actually this order of approximation gives the zero mass limit). We see
again that the large distance behavior remains unchanged and therefore
the skyrmion mass is not affected by the presence of light
fermions as well. A power law short distance behavior, however,
is introduced.

\section{The Introduction of Electromagnetic Coupling Between the Fermions}

\setcounter{equation}{0}
\bigskip

In the previous section, we analyzed the effects
produced on the skyrmion correlation function by the inclusion of fermions 
in the theory. The interaction among the fermions themselves, however, 
was neglected. This does not correspond to the realistic situation in any
possible application to a condensed matter system where the fermions are
electrons which, of course, feel their mutual electromagnetic interaction.
Therefore, let us consider in this section the case where the fermions,
in spite of moving on the plane possess a real electromagnetic interaction. 
The 2+1-dimensional theory which describes the {\it true} 
electromagnetic interaction in this case is not Maxwell but a modified 
version of it \cite{enp}. Taking into account the results of \cite{enp},
we can write the $CP^1$ lagrangian in the presence of fermions which
interact electromagnetically among themselves as
$$
\cl = 2 \ (D_\mu z_i)^\dagger (D^\mu z_i) + i \fr{\theta}{2}
\epsilon^{\mu\alpha\beta} A_\mu \del_\alpha A_\beta +
i \bar \psi \not\! \del \psi
- m \bar \psi \psi
-  \bar \psi \gamma^\mu \psi ( q A_\mu + e \ca_\mu ) + 
$$
\be
-\fr{1}{4} \cf_{\mu\nu} \lef [ \fr{1}{(-\Box)^{1/2}} \ri ] \cf^{\mu\nu}
\label{45}
\ee
where $e$ is the electric charge and $\ca_\mu$ is the
2+1D electromagnetic field.
Again, the integration over the $CP^1$-fields and also over
the fermion fields can be made as before. The integration over 
$\ca_\mu$, is then
quadratic and we now arrive at an
expression for the skyrmion correlation function which is identical to
(\ref{26}), except for the modifications
$$
\theta \longrightarrow \theta + A
$$
\be
M \longrightarrow M + B (-\Box) + \fr{e^2}{q^2} \lef[ B^2 (-\Box)^{3/2} +
 A^2 (-\Box)^{1/2} \ri ]
\label{46}
\ee
which is valid up to the order $e^2$.
$A$ and $B$ have been displayed in the previous section and are given,
respectively, in the large and small fermion
mass limit by (\ref{37}) and (\ref{41}). Observe that both of them are
of the order $O(q^2)$ and therefore the last term in the second piece of
(\ref{46}) is of the order $O(e^2q^2)$.

Inserting (\ref{46}) in (\ref{27}), (\ref{29})
and thereafter, we obtain an expression identical to (\ref{32}) except
for the fact that the last term between brackets is now modified as
\be
\lef [ \fr{M}{(-\Box)^2} \ri ] \longrightarrow
\lef [ \fr{M}{(-\Box)^2} + \fr{B}{(-\Box)}+
\fr{e^2}{q^2} \lef [ B^2  (-\Box)^{1/2}   + \fr{A^2}{(-\Box)^{3/2}} 
\ri ] \ri ]
\label{47}
\ee
Following the same steps as before and using (\ref{37}),
we obtain, in the large mass limit, up to the order $\fr{1}{m}$,
\be
<\mu(x) \mu^\dagger(y)>_{e,LM} = 
\fr{\exp \lef \{- \fr{ \rho_0^2 b^2}{2 \pi} |x - y| +
\fr{q^2 b^2}{64 \pi^2 m |x - y|}\ri \}}
{|x - y|^{\fr{e^2 q^2 b^2}{ 16 \pi^2}}}
\label{48}
\ee

In the small mass limit, we go through the same steps but use (\ref{41})
for $A$ and $B$ in (\ref{46}). The result is
$$
<\mu(x) \mu^\dagger(y)>_{e,SM} = 
\exp \lef \{- \fr{\rho_0^2 b^2}{2 \pi} |x - y| -
\fr{q^2 b^2}{ 32 \pi^2} \lef [ 1 +
e^2 \lef (\fr{ \pi^2 + 16}{ 16 \pi^2}\ri ) \ri ] \ln |x - y| \ri \} =
$$
\be
= \fr{\exp \lef \{- \fr{ \rho_0^2 b^2}{2 \pi} |x - y|
\ri \}}{|x - y|^{\fr{q^2 b^2}{ 32 \pi^2}
\lef [ 1 + e^2 \lef (\fr{ \pi^2 + 16}{ 16 \pi^2}\ri ) \ri ] }}
\label{49}
\ee
Observe that in the limit $e \rightarrow 0$, (\ref{48}) and (\ref{49})
reduce to (\ref{40}) and (\ref{44}), respectively.

\section{The Statistics of Skyrmions}

\setcounter{equation}{0}
\bigskip

It is well known that in the presence of a Chern-Simons term, skyrmions
acquire a generalized statistics determined by $\theta$ \cite{ss}. We can
see this fact in the present formulation in a quite interesting way. The
statistics of a particle implies certain commutation relation of
the quantized operator associated to it. When we evaluate the euclidean
correlation functions of these field operators, the commutation relation
manifest itself as a multivaluedness of these euclidean correlators
\cite{es,evora}, each sheet of the function being associated with a certain
ordering of operators. In the bosonic case, of course, the correlators are
univalent. Consequently, we can easily infer the statistics of skyrmion
in the different cases analyzed above, by checking the multivaluedness
of the corresponding correlation functions.

As remarked at the end of Sect. 3 there is a cancellation between 
$S_{2,L}$, given in (\ref{31}) and $T_2$, given by (\ref{33}) in all of
the correlation functions evaluated in Sects. 3,4,5. The only difference
in each case is the value of $\theta$, which in the latter cases is modified
according to (\ref{38}), (\ref{42}) or (\ref{46}). As can be clearly seen
from (\ref{33}), however, either $T_2$ or $S_{2,L}$ are defined up to a
factor
\be
i \varphi = i\ \  n\  \lef ( \fr{\theta b^2}{4 \pi} \ri ) 2 \pi
\label{499}
\ee
where $n=0,\pm 1,...$. We are therefore led to the conclusion that the
correlation functions (\ref{34}), (\ref{40}), (\ref{44}), (\ref{48}) and
(\ref{49}) are multivalued and defined up to a phase $\varphi$, given by
the above expression. The value of $\theta$ changes according to
(\ref{38}), (\ref{42}) or (\ref{46}), in each case. The skyrmion statistics,
in the pure CP$^1$-NSM is, therefore,
\be
S = \fr{\theta b^2}{4 \pi}
\label{50}
\ee
This is in agreement with the result found in \cite{ss}.
When we couple the model to fermions we see from
(\ref{38}), (\ref{42}) and (\ref{46}) that the modification of skyrmion
statistics is the same either in the presence of the electromagnetic
interaction or not, giving
\be
S_{LM} = \fr{\lef (\theta + \fr{q^2}{2 \pi}\ri ) b^2}{4 \pi}
\label{51}
\ee
and
\be
S_{SM} = \fr{\lef (\theta + \fr{q^2}{4 \pi}\ri ) b^2}{4 \pi}
\label{52}
\ee
respectively in the large and small mass limits.

\section{Conclusion and Remarks}

In this work, the correlation functions of quantum skyrmion fields were
explicitly evaluated in three different physical situations involving the
2+1-dimensional O(3) Nonlinear Sigma Model in its CP$^1$ version. The
relevant functional integrals were evaluated through a saddle-point method
but full quantum effects are described by the use of skyrmion creation
operators .
The mass and statistics of the skyrmion quantum excitations were
obtained by an analysis of the behavior of these correlation functions.
Our results show that the inclusion of fermions, either in the presence of
an electromagnetic interaction or not, does not change the skyrmion mass
which is obtained in the pure NSM case. This is shown to be smaller than
the classical one by a factor 0.5, confirming the tendency of quantum
effects to reduce the mass, which was observed at
semiclassical level \cite{rod}.
The statistics of skyrmions is changed due to the presence of fermions
but is not sensible to presence of an electromagnetic interaction. The
short distance behavior of the skyrmion correlation functions and therefore
the ultraviolet properties, on the other hand,
are completely changed, both by the mere introduction of fermions and
by turning on the electromagnetic interaction among them. It is a well
known fact in field theory that the existence of asymptotic states associated
to a certain quantum field requires a singular short distance behavior of the
corresponding correlation function. Comparing (\ref{34}) with
(\ref{40}), (\ref{44}), (\ref{48}) and (\ref{49}), we conclude that only in
the presence of fermions we expect asymptotic quantum skyrmion states. This
may be related to the so called marginal stability of the classical
skyrmion solution in the pure theory, which is associated to
a scale invariance which may be broken by the introduction of $q, e$ and
$m$. This result is rather suggestive, as several authors demonstrated
that real skyrmions are introduced in the system through electron doping
\cite{st,rod,go,sh,emsc}.
Looking at the above mentioned expressions obtained
for the skyrmion correlation function, on the other hand,
we see that always $<\mu> =0$,
implying that the skyrmions never condense, contrary to what was conjectured
in \cite{fm}, for $\theta=0$.

There is a great potential of applicability of the
present formalism in planar
condensed matter systems which can be described by the NSM in the continuum
limit. These include high temperature superconductors and systems presenting
the quantum Hall effect. A preliminary application in the first case has
been made in \cite{emsc}. In order to describe the doping process, a
constraint must be put on the fermion density. This would affect the
skyrmion mass in an important way, as described in \cite{emsc}.
Another interesting application we can envisage for the present
formalism
is the calculation of magnetic form factors in the presence of skyrmions
using the quantum skyrmion fields and correlation functions. This could
have measurable consequences in neutron scattering experiments.
The study of the recently experimentally observed
skyrmions in quantum Hall systems \cite{sk}, on the other hand,
is now being performed
using the formulation introduced in this paper.
The nonrelativistic version of the CP$^1$ model, however, must be
considered in this case. In conclusion, we stress that there are
many possibilities of extracting interesting results from the formulation
introduced here in real condensed matter systems. We are presently
exploring some of them.


\begin{thebibliography}{99}



\bibitem{bp} A.A.Belavin and A.M.Polyakov, {\it JETP Lett.} {\bf 22},
245 (1975)

\bibitem{zhk} S.C.Zhang, T.Hansson and S.Kivelson, 
{\it Phys. Rev. Lett.} {\bf 62}, 8 (1989);
N.Read, {\it Phys. Rev. Lett.} {\bf 62}, 86 (1989)

\bibitem{skk} D.H.Lee and C.L.Kane,
{\it Phys. Rev. Lett.} {\bf 64}, 1313 (1990);
S.L.Sondhi, A.Karlhede and S.Kivelson,
{\it Phys. Rev.} {\bf B 47}, 16416 (1993)

\bibitem{sk} S.E.Barnett et al. {\it Phys. Rev. Lett.} {\bf 74}, 5112 (1995);
A.Schmeller et al. {\it Phys. Rev. Lett.} {\bf 75}, 4290 (1995); 
E.H.Aifer et al. {\it Phys. Rev. Lett.} {\bf 76}, 680 (1996)

\bibitem{ha} F.D.M.Haldane, {\it Phys. Rev. Lett.} {\bf 50}, 1153 (1983);
{\it Phys. Lett.} {\bf A93}, 464 (1983)

\bibitem{dk} E.Dagotto, {\it Rev. Mod. Phys.} {\bf 66}, 763 (1994)

\bibitem{ak} A.P.Kampf, {\it Phys. Rep.} {\bf 249}, 219 (1994)

\bibitem{st} B.I.Schraiman and E.D.Siggia, {\it Phys. Rev. Lett.} {\bf 61},
467 (1988); {\it Phys. Rev.} {\bf B 42}, 2485 (1990) 

\bibitem{rod} J.P.Rodriguez, {\it Phys. Rev.} {\bf B39}, 2906 (1989);
{\bf B41}, 7326 (1990)

\bibitem{go} R.J.Gooding, {\it Phys. Rev. Lett.} {\bf 66}, 2266 (1991);
R.J.Gooding and A.Mailhot, {\it Phys. Rev.} {\bf B 48}, 6132 (1993)

\bibitem{sh} S.Haas, F.C.Zhang, F.Mila and T.M.Rice,
{\it Phys. Rev. Lett.} {\bf 77}, 3021 (1996)

\bibitem{emsc} E.C.Marino, {\it ``A Model for Doping in
High-Temperature Superconductors''}, {\it Phys. Lett.} {\bf A}, in press


\bibitem{pw} P.B.Wiegmann, {\it Phys. Rev. Lett.} {\bf 60}, 821 (1988)

\bibitem{fm} K.Furuya and E.C.Marino, {\it Phys. Rev.} {\bf D41}, 727 (1990)

\bibitem{nv} E.C.Marino, {\it Int. J. Mod. Phys.} {\bf A10}, 4311 (1995)

\bibitem{qms} E.C.Marino, {\it Phys. Rev.} {\bf D53}, 1001 (1996) 

\bibitem{vor} E.C.Marino, {\it Phys. Rev. } {\bf D38}, 3194 (1988)

\bibitem{enp} E.C.Marino, {\it Nucl. Phys.} {\bf B408} [FS], 551 (1993)

\bibitem{pol} A.M.Polyakov, {\it Gauge Fields and Strings}, Harwood,
New York (1987)

\bibitem{raj} R.Rajaraman, {\it Solitons and Instantons}, North Holland,
Amsterdam (1982)

\bibitem{ss} F.Wilczek and A.Zee, {\it Phys. Rev. Lett.}, 2250 {\bf 51}
(1983)

\bibitem{evora} E.C.Marino, {\it ``Dual Quantization of Solitons''}
in {\it Applications of Statistical
and Field Theory Methods in Condensed Matter}, D.Baeriswyl, A.Bishop and
J.Carmelo, eds., Plenum, New York (1990)


\bibitem{cmfc} E.C.Marino, {\it Phys. Rev.} {\bf D55}, 5234 (1997)

\bibitem{fd} A.Coste and M.L\"uscher, {\it Nucl. Phys.}, 631 {\bf B323}
(1989); D.G.Barci, C.D.Fosco and L.E.Oxman, {\it Phys. Lett.}, 267
{\bf B375} (1996)


\bibitem{es} E.C.Marino and J.A.Swieca, {\it Nucl. Phys.} {\bf B170} [FS1],
175 (1980)



\end{thebibliography}
\end{document}